\documentclass{optica-article}

\journal{oe}
\usepackage{lineno}
\usepackage{bm, comment}
\usepackage{booktabs}
\usepackage{color}
\usepackage{adjustbox}

\newif\ifchanges
\newcommand{\change}[1]{\ifchanges \textcolor{red}{#1} \else #1 \fi}

\changesfalse

\begin{document}

\title{Deep sound-field denoiser: optically-measured sound-field denoising using deep neural network}

\author{Kenji Ishikawa,\authormark{1,*} Daiki Takeuchi,\authormark{1} Noboru Harada,\authormark{1} and Takehiro Moriya\authormark{1}}

\address{\authormark{1}NTT Communication Science Laboratories, Nippon Telegraph and Telephone Corporation, Atsugi, Kanagawa 243-0198, Japan}

\email{\authormark{*}ke.ishikawa@ntt.com} %% email address is required; see note below about the corresponding author designation

% \homepage{http:...} %% author's URL, if desired

%%%%%%%%%%%%%%%%%%% abstract %%%%%%%%%%%%%%%%
%% [use \begin{abstract*}...\end{abstract*} if exempt from copyright]

\begin{abstract*}
This paper proposes a deep sound-field denoiser, a deep neural network (DNN) based denoising of optically measured sound-field images. Sound-field imaging using optical methods has gained considerable attention due to its ability to achieve high-spatial-resolution imaging of acoustic phenomena that conventional acoustic sensors cannot accomplish. However, the optically measured sound-field images are often heavily contaminated by noise because of the low sensitivity of optical interferometric measurements to airborne sound. Here, we propose a DNN-based sound-field denoising method. Time-varying sound-field image sequences are decomposed into harmonic complex-amplitude images by using a time-directional Fourier transform. The complex images are converted into two-channel images consisting of real and imaginary parts and denoised by a nonlinear-activation-free network. The network is trained on a sound-field dataset obtained from numerical acoustic simulations with randomized parameters. We compared the method with conventional ones, such as image filters, a spatiotemporal filter, and other DNN architectures, on numerical and experimental data. The experimental data were measured by parallel phase-shifting interferometry and holographic speckle interferometry. The proposed deep sound-field denoiser significantly outperformed the conventional methods on both the numerical and experimental data. Code is available on GitHub: https://github.com/nttcslab/deep-sound-field-denoiser.
\end{abstract*}

%%%%%%%%%%%%%%%%%%%%%%%%%%  body  %%%%%%%%%%%%%%%%%%%%%%%%%%
\section{Introduction}
Optical imaging has recently been used for high-spatial-resolution imaging of acoustic phenomena in airborne sound fields that conventional acoustic sensors cannot accomplish~\cite{Ishikawa2016}. The acousto-optic effect~\cite{Torras2012}, in which the refractive index of a medium is changed by sound, allows sound to be measured from the optical phase variation. Various optical methods have been used, for example, laser Doppler vibrometry (LDV)~\cite{Oikawa2005, Torras2012}, parallel phase-shifting interferometry (PPSI)~\cite{Ishikawa2016}, and digital holography~\cite{Matoba2014, Takase2021, Rajput2021, Hassad2022}. The applications include investigation of acoustic phenomena~\cite{Ishikawa2018, Ishikawa2020, Tanigawa2020} and sound-field back-projection~\cite{Oikawa2005, Torras2012, Yatabe2017, Verburg2021, Verburg2022}. Owing to their significant advantages, optical technologies are considered promising as a next-generation acoustic sensing modality. 

The sound field measured by a high-speed camera or scanning laser beam can be represented as an image sequence. Each pixel value is proportional to the line integral of the sound pressure along the corresponding optical path with superimposed noise. Because the phase fluctuation of light caused by audible sound is tiny owing to its physical origin, noise reduction of sound-field images is a fundamental concern. \change{The noises involved in optical imaging of sound field can include electronic noises of an image sensor, optical noises such as shot noise and laser intensity and phase noises, speckle noises, and environmental noises such as seismic vibration and atmospheric turbulence. To reduce these various noises and increase the signal-to-noise ratio, it is necessary to denoise the sound-field data.} Whereas sound-field image denoising is typically conducted using filters, as discussed in section~\ref{subsec:conv}, no machine-learning-based sound-field image denoising method has been proposed so far.

\begin{figure*}[t]
	\centering
	\includegraphics[width= \linewidth]{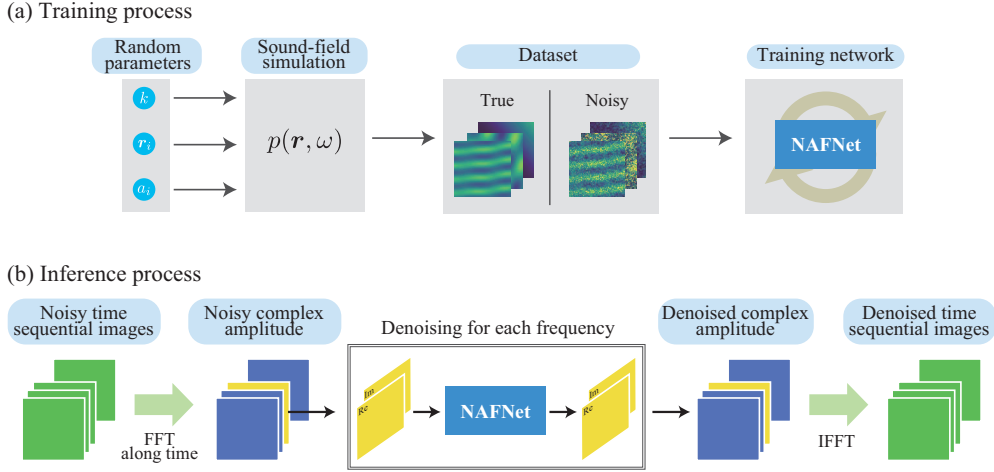}
	\caption{Overview of the deep sound-field denoiser. (a) Training process. A sound-field dataset is generated in a 2D acoustic simulation with randomized parameters. Each data is a complex-amplitude sound-field image of a harmonic frequency $\omega$. A nonlinear activation-free network (NAFNet) is trained using the clean and noisy pairs of the simulated sound fields. (b) Inference process. The time-sequential sound-field images are transformed into complex amplitude images, where each image is denoised by the trained network.}
	\label{fig:overview}
\end{figure*}

In this paper, we propose a denoising method for sound-field images that is based a deep neural network (DNN) (Fig.~\ref{fig:overview}). A sound-field image sequence is Fourier transformed along the time direction at each pixel, by which is obtained complex-amplitude sound-field images corresponding to the frequency bins of the discrete Fourier transform (FT). Then, each complex-amplitude image is converted into a two-channel image consisting of real and imaginary parts and denoised by using a trained DNN. To train the network, we generated training datasets by performing acoustic simulations with white and/or speckle noises. Randomizing the simulation parameters ensured variety in the training data. Numerical experiments confirmed that the proposed DNN-based method performs better than conventional methods. We also applied the method to data measured by PPSI and holographic speckle interferometry (HSI). It outperformed conventional methods on these experimental data without priori knowledge of the sound field.

\section{Related works}
\subsection{Sound-field denoising \label{subsec:conv}}
The physical properties of sound are commonly utilized for designing noise-reduction filters. These filters can be categorized into time-domain processing, spatial-domain processing, and spatiotemporal-frequency-domain processing.

Time-domain processing is typically the first choice. Because sound pressure varies over time, a high-pass filter with a very low cutoff frequency can eliminate static optical phase components and low-frequency fluctuations caused by air fluctuation and seismic vibration. Taking the difference between successive frames of an image sequence, which is a simple high pass filter, has been used as an easy denoising method for sound-field image sequences~\cite{Ishikawa2018}. When the frequencies of a measured sound field are known, the noise-reduction performance can be improved by designing an appropriate temporal filter~\cite{Yatabe2018}.

Spatial-domain processing can be applied independently of the time-domain processing. A spatial filter is applied to the sound-field image at each frame. Since sound is a spatially smooth variation and steep edges are usually absent, typical image processing filters, such as Gaussian and median filters, are effective~\cite{Chitanont2017}.

Spatiotemporal-frequency-domain processing utilizes the fact that sound satisfies the equation: $k = \omega / c$, where $k$ is the acoustic wavenumber, $\omega$ is the acoustic angular frequency, and $c$ is the speed of sound. If we consider a two-dimensional space, this equation forms a cone in $k-\omega$ space~\cite{Chitanont2017}. Since all of the spatiotemporal components in the recorded images that do not exist on the cone are noise, they can be eliminated by filtering ~\cite{Chitanont2017, Tanigawa2019}.

The methods used so far are all classical filters. We developed a DNN-based denoising of sound fields and confirmed that it outperforms these conventional methods.

\subsection{Natural image denoising by DNNs}
DNNs have been extensively applied to image-denoising tasks and have outperformed classical methods. Convolutional neural networks (CNN) ~\cite{unet2015, n2n2018, n2v2019, Yue2020, Zamir2021,nafnet2022} and transformers~\cite{Chen2021, swinir2021, Uformer2022, restormer2022} have widely been used. Among the numerous DNNs, a nonlinear activation free network (NAFNet)~\cite{nafnet2022} has a simple and efficient structure and has achieved  peak signal-to-noise ratio (PSNR) of 40.30 dB on a smartphone image denoising dataset~\cite{ssid2018}. We chose this architecture for our sound-field denoiser.

\subsection{DNNs for optical metrology}
DNNs have been increasingly used in optical metrology~\cite{Zuo2022}. DNNs have been used in many processes, including pre-processing (e.g., fringe denoising~\cite{Yan2019} and enhancement~\cite{Shi19}), analysis (e.g., phase retrieval~\cite{Feng2019} and phase unwrapping~\cite{Wang19}), and post-processing (e.g., phase denoising~\cite{Montresor2020}, error compensation~\cite{Nguyen2017}, and digital refocusing~\cite{Ren2018}). 

Several DNN-based methods have shown high performance in denoising fringe patterns and optical phase maps. For fringe denoising, a deep CNN consisting of 20 layers was proposed by Yan \emph{et al.}, where the training dataset was generated from Zernike polynomials and additive white Gaussian noise. Several methods have also applied DNNs to fringes corrupted by speckle noise~\cite{Jeon2018, Hao2019, Lin2020, Reyes2021, Wang2022, Javier2022}. Similar ideas have been used to denoise optical phase maps~\cite{Montresor2020, Yan2020a, Yan2020b, Li2021, Li2022, Fang2022}. 

However, no research has used DNNs to denoise sound-field images measured by optical methods. Since the spatial and temporal features of sound-field images differ from those of interference fringes and typical optical phase maps, the previous methods may not be optimal for sound-field denoising. Our contribution is that we developed DNN-based sound-field denoising methods and a training dataset that considers the physical nature of sound.

\section{Methods \label{sec:method}}

\subsection{Acousto-optic measurement data}
Here, let us briefly review the principle of acousto-optic measurement~\cite{Torras2012}. The acousto-optic effect is the change in the refractive index of a medium caused by sound. If light propagates along the $z$-axis, the phase shift of the light propagating through a sound field in air is given by
\begin{equation}
 \phi_s (x, y, t) = k_L \frac{n_0-1}{\gamma P_0} \int_{z_1}^{z_2} p(x, y, z, t) dz,
\end{equation}
where $k_L$ is the wavenumber of light, $\gamma$ is the specific heat ratio, $n_0$ and $P_0$ are the refractive index and pressure of air in a static condition, respectively, and $p$ is the sound pressure. \change{The typical values in air are $n_0 = 1.000279$, $\gamma = 1.40$ and $P_0 = 101325$, and  $k_L$ is calculated from wavelength of light.}
The phase shift of light is proportional to the sound pressure along the laser path. When sound-field imaging is performed based on this principle, the observed data can be written as a three-dimensional array $\Phi_\mathrm{noisy}$ whose elements are of the form $\phi_s (x_i, y_j, t_m)$ , where $(i, j)$ is the pixel index and $m$ is the time index. Any processing method that can extract $\phi_s$ from noisy data can be applied.

\subsection{DNN-based sound-field denoising}
The overview of the inference process is shown in Fig.~\ref{fig:overview}(b). 
First, \change{the temporal Fourier analysis is used to get the complex-valued amplitude at a certain acoustical frequency. A time-domain FT is calculated at each of all pixels of the time-sequential sound-field images. This can be written as } $\Psi_\mathrm{noisy} = \mathcal{F}_t[\Phi_\mathrm{noisy}]$, where $\mathcal{F}_t$ denotes a 1D FT along the temporal axis. $\Psi_\mathrm{noisy}$ is the complex-valued amplitude at the corresponding spatial position and the Fourier frequency. Then, for each Fourier frequency, the 2D complex-valued amplitude is converted into a two-channel image with real and imaginary parts. The two-channel complex-amplitude image is normalized and inputted to the neural network. The network is trained to output a clean complex-amplitude image from the input noisy complex-amplitude image. The output image is multiplied by the reciprocal of the normalization factor to maintain the magnitude of the sound field. After processing all frequencies independently with the same DNN, the denoised complex amplitude, $\Psi_\mathrm{denoise}$, is inverse Fourier transformed, and the denoised sound field $\Phi_\mathrm{denoise} = \mathcal{F}_t^{-1}[\Psi_\mathrm{denoise}]$ is obtained.

Since the proposed method uses DNNs to denoise two-channel input images, any network that is able to perform denoising on images can be used with it. Here, Unet-based networks are often used in optical metrology~\cite{Zuo2022}. In particular, we chose NAFNet~\cite{nafnet2022}, which has excellent performance and can run with relatively small memory and training time. %A discussion of the optimal network structure and parameters for denoising the sound field may be a subject of future work.

\begin{figure*}[t]
	\centering
	\includegraphics[width= \linewidth]{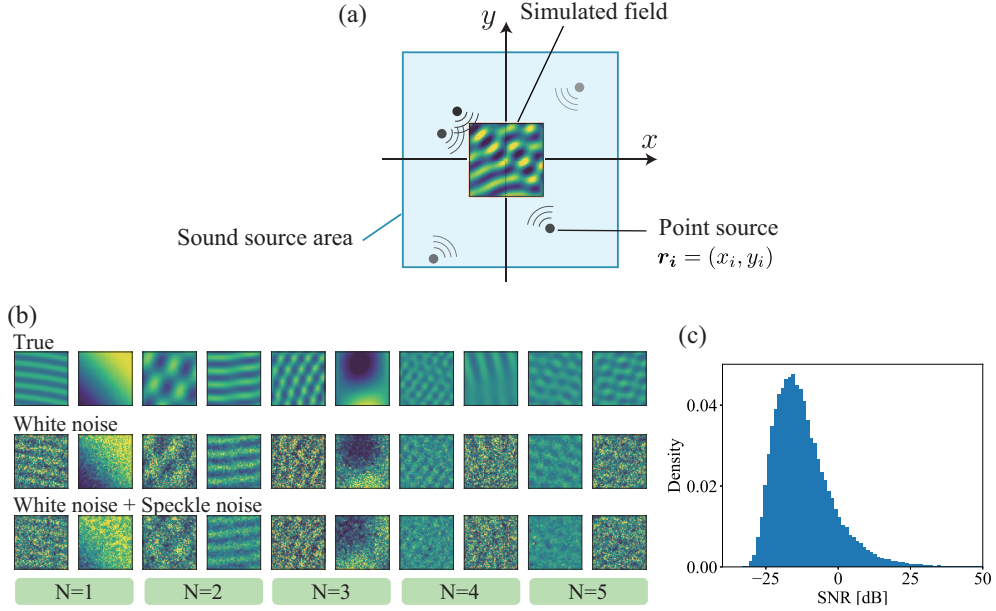}
	\caption{(a) Sound-field data generation. Point sources are randomly generated within the sound source area, and the 2D true sound fields in the center area are generated using the Green's function of the 2D Helmholtz equation. (b) Examples of the generated sound-field data. $N$ represents the number of sound sources. Two examples are shown for each $N$. 
 \change{(c) Histogram of SNR of the generated white noise images.} }
	\label{fig:data}
\end{figure*}

\subsection{Training data}
Although optical sound measurements have been actively studied in recent years, no dataset exists for training a neural network on them. It is difficult to collect sound-field data under various conditions through experiments. Therefore, this study used acoustic numerical simulation to create a training dataset\change{, as shown in Fig.~\ref{fig:overview}(a)}.

A 2D sound-field simulation with randomized parameters was used. \change{For generating the training dataset, we created the data as a 2D sound field instead of calculating line-integral of 3D sound field for reducing computational costs and complexity.} Figure~\ref{fig:data}(a) shows a schematic illustration of the simulation. The inner rectangle is the measurement area, outside of which is the sound source area where point sources are randomly placed. To generate sound fields with diverse spatial characteristics from simple to complex, the number of point sources was varied from 1 to 5, and the position and relative amplitude of each source was randomly assigned. Each true sound field is a superposition of the sound waves generated by these point sources and can be calculated as
\begin{equation}
 p_\mathrm{image, true} \left(\bm{r}, k\right) = A \sum^{N}_{i=1} a_i \frac{j}{4} H^{(2)}_0 \left(k |\bm{r_i} - \bm{r}| \right),
\end{equation}
where $\bm{r}=(x,y)$, $k$ is the magnitude of acoustic wavenumber, $A$ is a constant determining the overall magnitude of the sound field, $N$ is the number of sound sources, $a_i$ and $\bm{r_i}=(x_i, y_i)$ are the relative amplitude and position of the $i$th sound source, respectively, and $H^{(2)}_0$ is a Hankel function of the second kind of order zero. 
\change{The term $(j/4) H^{(2)}_0 \left(k |\bm{r_i} - \bm{r}| \right)$ is Green's function of a 2D wave equation, which describes a sound field created by a point source~\cite{FourierAcoustics}. Therefore, each term in the summation symbol represents a sound field by a point source of position $\bm{r_i}$ and amplitude $a_i$.}
% Inside $\Sigma$ is the product of the relative amplitude of the $i$th source and Green's function of the 2D Helmholtz equation.

The true sound fields were created by randomly selecting $k$, $a_i$, and $\bm{r}_i$ from uniform distributions. The measurement area was a square of side length 1, and the sound source area was ten times larger than that. The random parameters were generated from uniform distributions of \change{$0.1 \leq a_i \leq 1$}, $1.26 \leq k \leq 40.2$, $0.5 \leq |x_i| \leq 10$, and $0.5 \leq |y_i| \leq 10$. \change{In this range of wavenumbers, the shortest wavelength is 0.156 times the size of the imaging field of view, while the longest wavelength is 5 times the size of the imaging field of view.} The amplitude of the entire sound field was set to $A = 0.1$. These parameters were determined based on the authors’ experience with typical experimental conditions of this measurement technology. $a_1$ was set to 1 regardless of the number of sources to avoid all sources having small amplitudes. The simulated data was calculated by discretizing the measurement area into 128 $\times$ 128 pixels. The top row of Fig.~\ref{fig:data}(b) shows examples of the generated sound fields. It can be seen that the generated sound fields have different complexities, wavelengths, and directions of arrival.

Two types of noise were added to the training data: additive white Gaussian noise and speckle noise. 
\change{Due to the lack of knowledge regarding the measurement noise of complex-valued sound fields in the frequency domain, white noise was used to represent such noise.}
The amplitudes of the white noise were randomly selected from a uniform distribution between \change{0 to $0.1 A$. The histogram of the SNR of the 50,000 white noise images produced is shown in Fig.~\ref{fig:data}(c); the mean and standard deviation of the SNR were approximately -12.1 dB and 10 dB, respectively.}
The method of generating the speckle noise data is shown in Section 1 of \href{https://opticapublishing.figshare.com/s/26592e8ab4d2dee2e4eb}{Supplement 1}.
Examples of the noisy training data are shown in Fig.~\ref{fig:data}(b). Data with different amounts of noise were generated. Although the differences between white and speckle noise may be difficult to recognize, spatially correlated random patterns appear in the speckle noise images. Such speckle noise can occur, for example, in a sound field observation using electronic speckle pattern interferometry and a holographic interferometer equipped with Fresnel lenses~\cite{Ishikawa2022}.

\subsection{Implementation details}
This study used almost the same network as in the original NAFNet article~\cite{nafnet2022}, except for the number of image channels. The network consisted of 32 blocks with widths of 32, two image channels (real and imaginary), and a 128 $\times$ 128 pixel image size. The root mean square error was used as the loss, Adam was used as the optimizer. \change{For the white noise dataset, the initial learning rate was set to 1e-3, while for the speckle noise dataset, it was set to 2e-4. The learning rate decreased exponentially by a factor of 0.95 per epoch.} A total of \change{50,000} training data were created, \change{10,000} for each number of sources. The training batch size was 32, and the epoch was 50. \change{It took approximately 10 hours to complete the training process using a single NVIDIA RTX A4000 GPU.} The network trained on the white noise dataset is denoted by Ours (W), and the one trained on the speckle noise dataset is denoted by Ours (W+S). The data for evaluation consisted of \change{2,500} sound fields (\change{500} for each number of sound sources) generated by simulation under the same conditions as those used for generating the training data. The peak signal-to-noise ratio (PSNR) and structural similarity (SSIM) were used as evaluation metrics.

\subsection{Conventional methods}
\change{Six} conventional denoising methods were used for comparison with the proposed method, i.e., 2D Gaussian filter, 2D median filter, \change{non-local means (NLM)~\cite{NLM}, block-matching and 3D filtering (BM3D)~\cite{BM3D}, windowed Fourier filtering (WFF)~\cite{Kemao2004}}, and spatiotemporal band pass filter (ST BPF). The kernel sizes for the Gaussian and median filters were set to 7 pixels. \change{NLM and BM3D used an estimated standard deviation from a noisy image. For NLM, the filtering parameter was experimentally determined to be twice the estimated standard deviation with the patch size of $7 \times 7$ and search area of $23 \times 23$. For BM3D, the default parameters of the Python package \textit{bm3d} were used.} The four filters were applied to the real and imaginary parts of the complex amplitude image, respectively. \change{WFF has demonstrated its superior performances for denoising speckle noise~\cite{Montresor2016}. WFF processed complex-valued images directly without splitting them into real and imaginary parts. The filtering threshold was set to three times the standard deviation of each image, as suggested in \cite{Kemao2005}.}

The ST BPF is a spatial frequency filter based on the wave equation~\cite{Chitanont2017}. In the wavenumber spectrum, sound components lie on the circumference of $k = (k_x^2+k_y^2)^{1/2} = \omega/c$. Therefore, noise can be reduced by removing the spatial frequencies that do not satisfy the equation. First, if the input signal is a time series of images, a 2D complex sound field for each frequency is obtained by taking a 1D FT. Next, a fourth-order image Butterworth band-pass filter was created for and applied to each 2D complex-amplitude image. The lower cutoff frequency was set to $0.5 k$, and the higher cutoff frequency was $1.2 k$, where $k$ is determined by the center frequency of each Fourier frequency bin. Note that since the image resolution was not very high, the bandwidth of the bandpass filter was set wide in order to avoid removing the broadened components by low-resolution 2D FT. The lower cutoff frequency was determined carefully to avoid erasing too many components near the origin in the wavenumber spectrum.

\change{In addition, for comparison with the existing DNN-based method used for optical fringe denoising, the two DNN methods, DnCNN~\cite{Zhang2017} and LRDUNet~\cite{Javier2022}, were employed. DnCNN was originally proposed for natural image denoising~\cite{Zhang2017} and subsequently applied for optical fringe denoising~\cite{Yan2019,Montresor2020}. We have followed the original network architecture by Zhang et al. with 20 layers. The initial learning rate was set to 1e-4 with exponential decay by a factor of 0.95. LRDUNet was recently proposed by Javier et al. and showed superior performances for fringe denoising compared with conventional DnCNN and U-net~\cite{Javier2022}. The same network architecture as Javier's paper was used. The initial learning rate was set to 1e-3 and decayed as for DnCNN. DnCNN and LRDUNet were trained by the same sound-field datasets with the proposed method. The training epoch was 50 for both networks.}

\begin{table}[tb]
\centering
\caption{PSNR and SSIM of denoising results for white noise data; \change{values are averages over the test dataset.}}
\label{tab:sim_white}
\scriptsize
\begin{tabular}{ccccccccccccccc}
\toprule
\multicolumn{1}{c}{} & \multicolumn{6}{c}{PSNR [dB]} & \multicolumn{6}{c}{SSIM} \\
\cmidrule(rl){2-7} \cmidrule(rl){8-13} 
$N$ & 1 & 2 & 3 & 4 & 5 & ALL & 1 & 2 & 3 & 4 & 5 & ALL \\ 
 \midrule 
 Noisy & 4.40 & 1.08 & 0.36 & -0.89 & -2.06 & 0.58 & 0.166 & 0.115 & 0.100 & 0.092 & 0.086 & 0.112 \\
 Gaussian & 18.6 & 19.3 & 20.2 & 20.0 & 19.5 & 19.5 & 0.646 & 0.646 & 0.667 & 0.658 & 0.647 & 0.653 \\
 Median & 18.2 & 15.4 & 14.8 & 13.5 & 12.4 & 14.8 & 0.540 & 0.407 & 0.363 & 0.320 & 0.285 & 0.383 \\
 \change{NLM} &  20.2 & 19.7 & 20.4 & 20.0 & 19.5 & 19.9 & 0.642 & 0.575 & 0.581 & 0.544 & 0.522 & 0.573\\
 \change{BM3D} & 22.4 & 20.2 & 20.2 & 19.7 & 19.1 & 20.3 & 0.709 & 0.617 & 0.594 & 0.571 & 0.554 & 0.609\\
 \change{WFF} & 20.4 & 19.2 & 19.2 & 18.9 & 18.5 & 19.2 & 0.852 & 0.711 & 0.646 & 0.603 & 0.568 & 0.676 \\
 ST BPF & 16.5 & 17.3 & 17.8 & 17.8 & 17.9 & 17.5 & 0.768 & 0.766 & 0.758 & 0.743 & 0.729 & 0.753 \\
 \midrule
 \change{DnCNN (W)} & -0.66 & -1.11 & -1.48 & -1.66 & -2.14 & -1.41 & 0.196 & 0.185 & 0.182 & 0.178 & 0.174 & 0.183\\
 \change{LRDUNet (W)} & 36.4 & 30.0 & 28.3 & 26.7 & 25.3 & 29.4 & 0.990 & 0.952 & 0.915 & 0.870 & 0.828 & 0.911\\
 Ours (W) & \textbf{40.1} & \textbf{33.9} & \textbf{31.8} & \textbf{29.8} & \textbf{28.0} & \textbf{32.7} & \textbf{0.995} & \textbf{0.980} & \textbf{0.964} & \textbf{0.939} & \textbf{0.907} & \textbf{0.957}\\
 \midrule
 \change{DnCNN (W+S)} & -0.97 & -1.61 & -1.98 & -2.15 & -2.58 & -1.86 & 0.150 & 0.164 & 0.166 & 0.166 & 0.163 & 0.162 \\
 \change{LRDUNet (W+S)} & 13.8 & 14.8 & 15.4 & 15.7 & 15.8 & 15.1 & 0.555 & 0.607 & 0.626 & 0.615 & 0.600 & 0.601\\
 Ours (W+S) & 13.7 & 15.0 & 15.9 & 16.3 & 16.5 & 15.5 & 0.555 & 0.623 & 0.654 & 0.653 & 0.642 & 0.625 \\
\bottomrule
\end{tabular}
\end{table}

\begin{figure}[t]
	\centering
	\includegraphics[width= \linewidth]{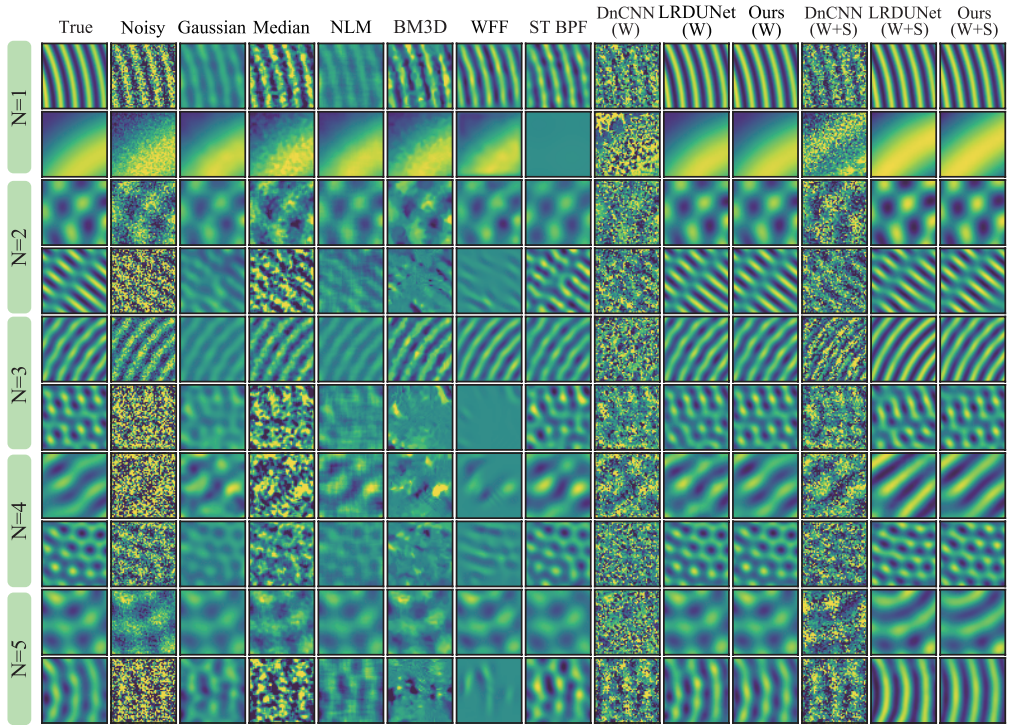}
	\caption{Examples of denoised images for white-noise data. Two examples are shown for each $N$. \change{The color bar range for all images is from -1 to 1.}}
	\label{fig:sim_white}
\end{figure}

\begin{figure}[t]
	\centering
	\includegraphics[width= \linewidth]{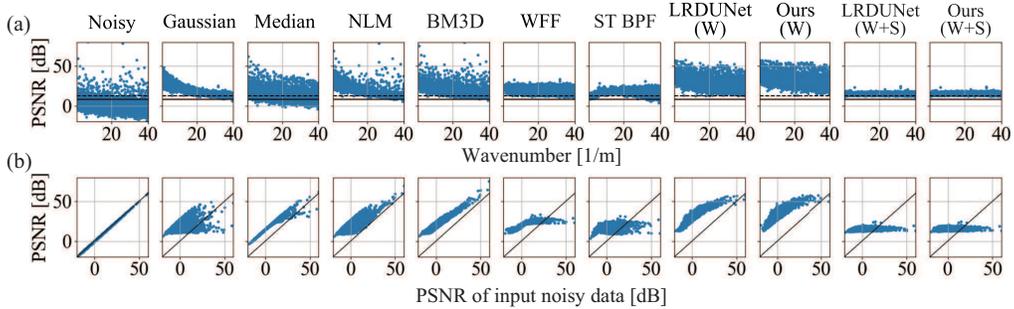}
    \caption{\change{PSNR plotted as a function of (a) acoustic wavenumber and (b) input noisy data for white-noise data. Solid and dashed lines in (a) indicate the averaged PSNRs of the two baselines; the solid line is for random data from a uniform distribution on the interval (-1, 1), and the dashed line is data with all zeros. }}	
 % \caption{\change{(a) PSNR plotted as a function of acoustic wavenumber and (b) Improvement of PSNR plotted as a function of PSNR of input noisy data for white-noise data. Solid and dashed lines in (a) indicate the averaged PSNRs of the two baselines. The solid line is for random data from a uniform distribution on the interval (-1, 1), and the dashed line is data with all zeros.}}
	\label{fig:res_sim_wh_scat}
\end{figure}

\section{Numerical results}
\subsection{Denoising of white noise data}
Table \ref{tab:sim_white} and Fig.~\ref{fig:sim_white} show the evaluation metrics and denoised sound-field images of the conventional and proposed methods for the white-noise data. The table shows that Ours (W) scored the highest in terms of PSNR and SSIM for all $N$ \change{included in the training data. In Section 2 of \href{https://opticapublishing.figshare.com/s/26592e8ab4d2dee2e4eb}{Supplement 1}, generalization results of Ours (W) for $N \in \{6, 7, 8, 9, 10 \}$ are shown.} Among the conventional methods, \change{BM3D} had the highest PSNR for the overall score, and ST BPF had the highest SSIM. Figure~\ref{fig:sim_white} shows that the Gaussian filter smoothed the noisy wavefront, but it blurred short-wavelength sound waves. The median filter performed worse than the other methods. \change{NLM, BM3D, and WFF restored sound fields relatively well when the noise was not severe, whereas they tended to lose almost all sound field information when the noise was significant, particularly in the fourth, sixth, seventh, and tenth rows.}
The ST BPF showed good overall results \change{except for very low wavenumber, as shown in the second row. When the noise was significant, the denoised sound fields exhibited noticeably different patterns from the true data because noises had spatial scales equivalent to the wavelength of sound passed the spatiotemporal filters. The DnCNN model failed to learn an appropriate mapping from the noisy data to clean data. LRDUNet (W) and Ours (W)} produced better noise reduction results than the conventional methods did, regardless of the sound-field parameters, such as the number of sound sources and acoustic wavelength, and the amount of noise. \change{LRDUNet (W+S) and Ours (W+S)} seemed to properly restore the wavefronts; nevertheless, its scores were significantly lower than those of LRDUNet (W) and Ours (W). \change{Comparing them with true data, it is evident that both LRDUNet (W+S) and Ours (W+S) simplified the sound field too much. This may be because these models were trained to eliminate speckle noise from the dataset, which resulted in oversimplifying the data lacking speckle noise.}
%one can see that the sound field structures are the same, but the overall amplitude of the image is larger in Ours(W+S). Such a pattern may be acquired during the training by the data containing speckle noise.}

\change{
Figure~\ref{fig:res_sim_wh_scat} plots the PSNRs of each denoising method, except for the DnCNN, on the 2,500 test data as a function of (a) wavenumber and (b) PSNR of input noisy data. From Fig.~\ref{fig:res_sim_wh_scat} (a), the dependence of the denoising performance on the wavenumber can be found. The Gaussian filter, NLM, and BM3D performed well for the low wavenumbers, but their performance deteriorated as the wavenumber increased; the PSNR values approach the baselines, random data and data with all zeros, shown by solid and dashed lines. The ST BPF had low scores for very low wavenumbers because the spatial frequency bandpass filter unintentionally eliminated the very low wavenumber components. Most of the data points of LRDUNet (W+S) and Ours (W+S) are aligned with the baseline. 
}

\change{
Figure~\ref{fig:res_sim_wh_scat} (b) indicates the change in PSNR values before and after denoising. As in the noisy data plot, if PSNRs do not change by denoising, the data points align on the straight line of slope one, as shown by the black solid line, and if the PSNR values improve, the data points are plotted above the line. Gaussian filter, WFF, and ST BPF show a decrease in PSNR when the input noisy data has a high PSNR, i.e., clean data. This suggests that some signal components are inadvertently removed during the processing, leading to a decrease in PSNR. In contrast, NLM, BM3D, LRDUNet (W), and Ours (W) show improvements in most conditions. Particularly, LRDUNet (W) and Ours (W) demonstrate substantial enhancements, highlighting the effectiveness of learning-based methods.
}

\subsection{Denoising of speckle noise data}
\begin{figure}[t]
	\centering
	\includegraphics[width= \linewidth]{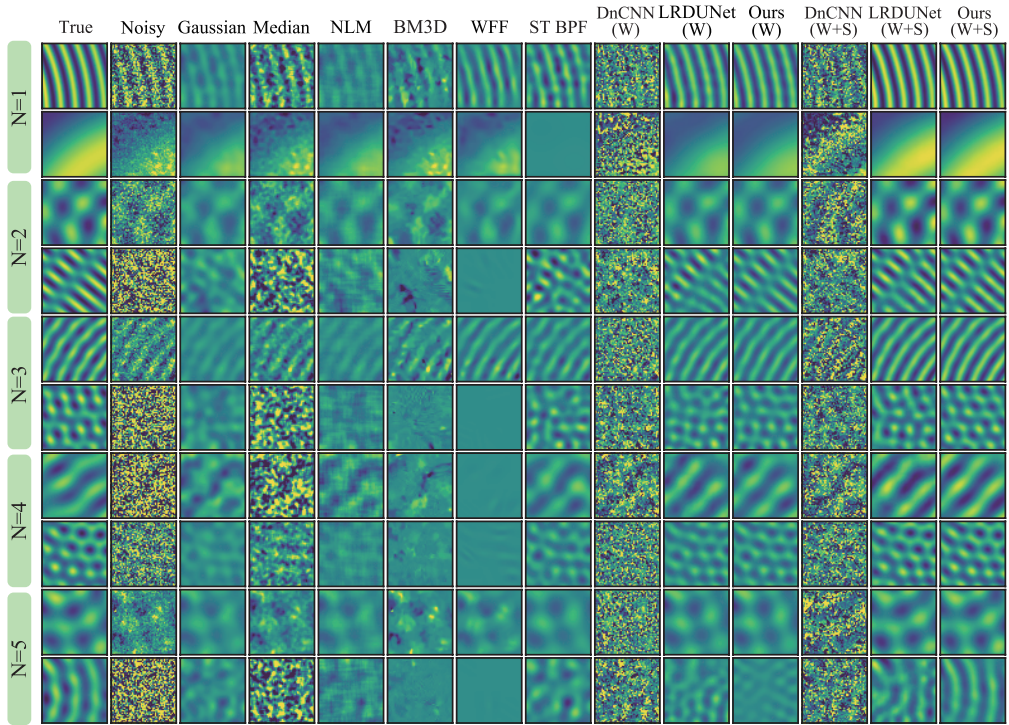}
	\caption{Examples of denoised images for the data with white and speckle noises. Two examples are shown for each $N$. \change{The color bar range for all images is from -1 to 1.}}
	\label{fig:res_sim_sp}
\end{figure}

\begin{figure}[t]
	\centering
	\includegraphics[width= \linewidth]{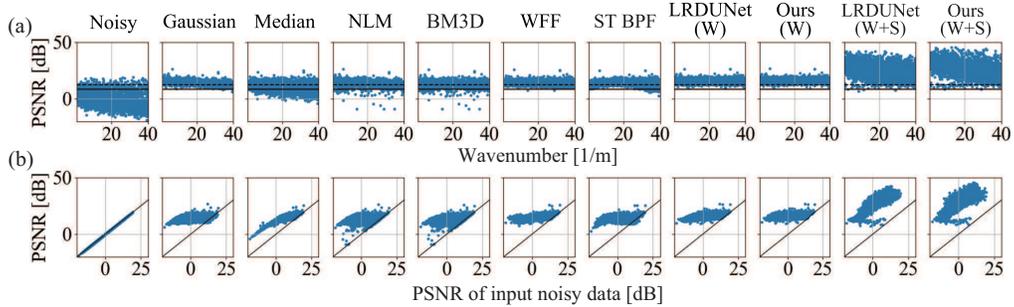}
	\caption{\change{PSNR plotted as a function of (a) acoustic wavenumber and (b) input noisy data for speckle noise data. Solid and dashed lines in (a) indicate the averaged PSNRs of the two baselines; the solid line is for random data from a uniform distribution on the interval (-1, 1), and the dashed line is data with all zeros.}}
	\label{fig:res_sim_sp_scat}
\end{figure}

\change{
Table \ref{tab:sim_speckle} and Fig.~\ref{fig:res_sim_sp} show the evaluation metrics and denoised sound-field images for the speckle noise data. The conventional filters, LRDUNet (W), and Ours (W) scored lower compared with their white-noise results, while LRDUNet (W+S) and Ours (W+S) scored higher. Figure~\ref{fig:res_sim_sp} indicates that conventional filters, especially when dealing with considerable noise, are challenging to restore the sound fields accurately. In addition, the networks trained on the white noise dataset may appear to perform well, but there are observed changes in the shape of the wavefront and a reduction in the magnitude. By contrast, LRDUNet (W+S) and Ours (W+S) significantly removed the noise and restored sound fields, except for LRDUNet (W+S) in the tenth row. The scatter plots of the PSNRs are shown in Fig.~\ref{fig:res_sim_sp_scat}. The PSNRs of the conventional filters, LRDUNet (W), and Ours (W) leveled off around 20 dB for almost all wave numbers, close to the zero data baseline. LRDUNet (W+S) and Ours (W+S) showed significant improvement for most of the data regardless of the wavenumber. These results confirm that the network properly learned the nonlinear transformation caused by speckle noise from the created training dataset.
}

\begin{table}[tb]
\centering
\caption{PSNR and SSIM of denoising results for data with white and speckle noises; \change{values are averages over the test dataset.}}
\label{tab:sim_speckle}
\scriptsize
\begin{tabular}{ccccccccccccccc}
\toprule
\multicolumn{1}{c}{} & \multicolumn{6}{c}{PSNR [dB]} & \multicolumn{6}{c}{SSIM} \\
\cmidrule(rl){2-7} \cmidrule(rl){8-13}
$N$ & 1 & 2 & 3 & 4 & 5 & ALL & 1 & 2 & 3 & 4 & 5 & ALL \\
 \midrule 
 Noisy & 2.0 & -0.35 & -0.77 & -1.75 & -2.85 & -0.75 & 0.316 & 0.234 & 0.224 & 0.175 & 0.164 & 0.223 \\
 Gaussian & 12.2 & 13.8 & 14.8 & 15.2 & 15.4 & 14.3 & 0.391 & 0.436 & 0.471 & 0.474 & 0.475 & 0.449 \\
 Median & 11.1 & 10.9 & 10.9 & 10.5 & 9.80 & 10.7 & 0.331 & 0.276 & 0.258 & 0.234 & 0.213 & 0.263 \\
 \change{NLM} & 11.6 & 13.1 & 14.1 & 14.7 & 14.8 & 13.7 & 0.506 & 0.462 & 0.476 & 0.436 & 0.423 & 0.460 \\
 \change{BM3D} & 11.4 & 12.8 & 13.7 & 14.4 & 14.5 & 13.4 & 0.630 & 0.584 & 0.599 & 0.561 & 0.562 & 0.587 \\
 \change{WFF} & 12.8 & 13.8 & 14.7 & 15.1 & 15.4 & 14.4 & 0.463 & 0.430 & 0.429 & 0.430 & 0.421 & 0.435\\
 ST BPF & 12.3 & 13.7 & 14.5 & 14.8 & 15.1 & 14.1 & 0.476 & 0.528 & 0.544 & 0.545 & 0.543 & 0.527\\
 \midrule
 \change{DnCNN (W)} & -0.85 & -1.36 & -1.7 & -1.83 & -2.34 & -1.62 & 0.316 & 0.276 & 0.266 & 0.238 & 0.236 & 0.267 \\
 \change{LRDUNet (W)} & 14.1 & 15.6 & 16.3 & 16.5 & 16.5 & 15.8 & 0.580 & 0.622 & 0.619 & 0.593 & 0.568 & 0.596 \\
 Ours (W) & 14.2 & 16.0 & 16.9 & 17.2 & 17.3 & 16.3 & 0.580 & 0.653 & 0.666 & 0.649 & 0.631 & 0.636 \\
 \midrule
 \change{DnCNN (W+S)} & -1.53 & -2.07 & -2.49 & -2.56 & -2.91 & -2.31 & 0.425 & 0.359 & 0.353 & 0.304 & 0.293 & 0.347 \\
 \change{LRDUNet (W+S)} & 31.2 & 24.6 & 23.4 & 22.2 & 21.1 & 24.5 & 0.963 & 0.872 & 0.821 & 0.757 & 0.707 & 0.824 \\
 Ours (W+S) & \textbf{33.8} & \textbf{27.3} & \textbf{25.6} & \textbf{24.0} & \textbf{22.7} & \textbf{26.7} & \textbf{0.974} & \textbf{0.923} & \textbf{0.886} & \textbf{0.826} & \textbf{0.779} & \textbf{0.877}\\
\bottomrule
\end{tabular}
\end{table}

\subsection{Denoising of transient and broadband sound field}
\begin{figure}[t]
	\centering
	\includegraphics[width=\linewidth]{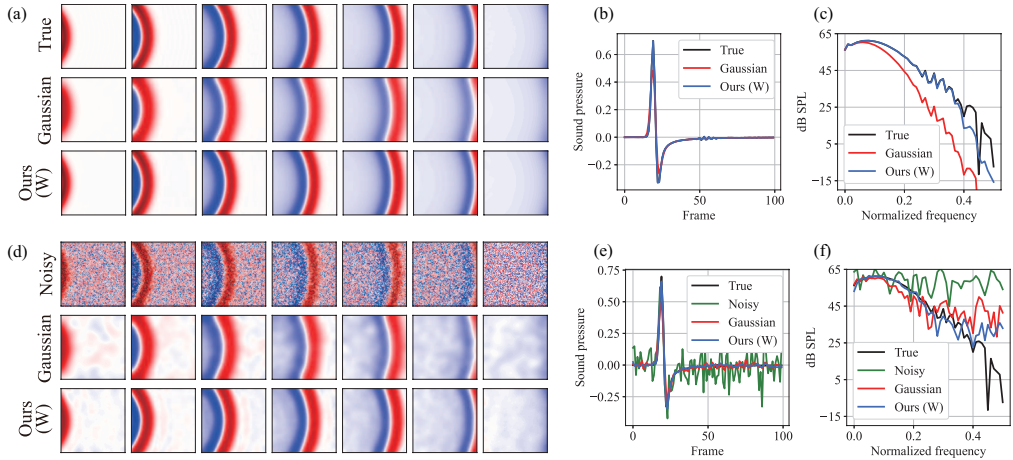}
	\caption{\change{Denoising of the transient sound field of the Gaussian pulse propagation. (a) True images and denoised images from the true data, (b) and (c) temporal signals and power spectra extracted from the center 4 $\times$ 4 pixels of the images in (a). (d) Noisy data and denoised images from the noisy data, (e) and (f) temporal signals and power spectra extracted from the center 4 $\times$ 4 pixels of the images in (d). The frames shown in (a) and (d) are from left to right: 10, 14, 18, 22, 26, 30, and 34.}}
	\label{fig:res_sim_tr}
\end{figure}

\change{
To verify the validity of the proposed method for transient and broadband signals, it was necessary to create and test sound field data in the time domain. The simulation of the sound field was conducted using COMSOL multiphysics to simulate a transient Gaussian pulse in a free field. The created sound field is shown in the top row of Fig.~\ref{fig:res_sim_tr}(a). Note that the Gaussian pulse is a transient and broadband signal because the Fourier transform of Gaussian shape is also Gaussian in the frequency domain.
}

\change{
Figure~\ref{fig:res_sim_tr}(a) also shows the denoising results of a noiseless Gaussian pulse sound field by the Gaussian filter and Ours (W). As no noise is present in the data, the denoising results should be identical to the True data. The Gaussian filter causes the wavefront's rise to become blurry, whereas Ours (W) remains almost identical to the True data. The waveforms extracted as an average of 4 $\times$ 4 pixels in the center of the image and their power spectra are shown in Fig.~\ref{fig:res_sim_tr} (b) and (c). As frequency increases, the deviation from the True data becomes more significant in the Gaussian filter, while Ours (W) maintains almost the same agreement with the True data in most frequency bands. 
}

\change{
Figure~\ref{fig:res_sim_tr}(d)-(f) shows the results for noisy data with white noise. As shown in Fig.~\ref{fig:res_sim_tr}(d), Ours (W) removes the noise and restores the original sound field. The waveforms in Fig.~\ref{fig:res_sim_tr}(e) ensure that the temporal information is preserved after denoising by the DNN. Furthermore, from the spectra in Fig.~\ref{fig:res_sim_tr}(f), one can find that Ours (W) restores the frequency spectrum of the clean signal buried in the white noise.  
These analyses indicate that the proposed deep sound-field denoiser reduces the noise in the transient and broadband sound field effectively and does not lose much information about the original sound field.
}

\color{black}

\section{Experiments}
We denoised experimental data measured by two optical systems: PPSI~\cite{Ishikawa2016}, in which the primary noise source was white noise, and HSI using Fresnel lenses~\cite{Ishikawa2022}, in which speckle noise was superimposed. 

\subsection{Parallel phase-shifting interferometry}
\begin{figure}[t]
	\centering
	\includegraphics[width= .9\linewidth]{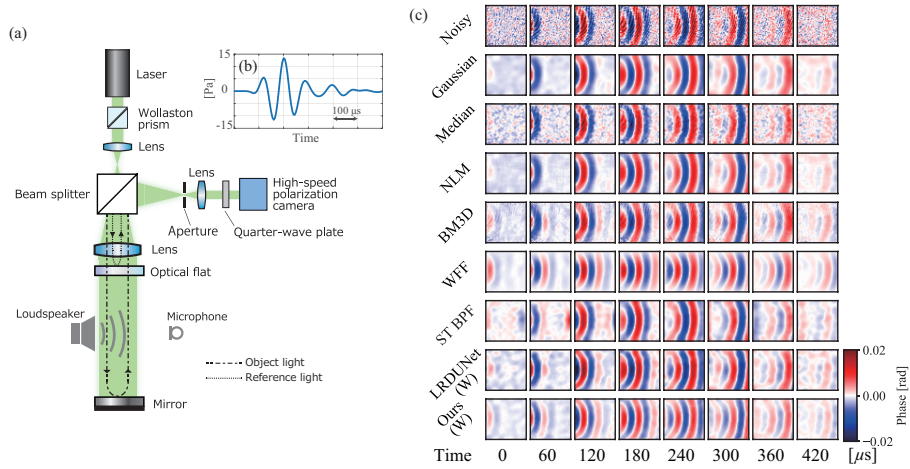}
	\caption{(a) Schematic diagram of the PPSI measurement system. A three-cycle burst wave with a center frequency of 12 kHz was emitted from the loudspeaker. (b) Sound pressure waveform measured by the microphone placed 20 cm from the loudspeaker's diaphragm. (c) Denoising results of transient sound fields measured by PPSI.}
	\label{fig:ppsi}
\end{figure}

PPSI is a system that combines a Fizeau interferometer and a polarized high-speed camera, as shown in Fig.~\ref{fig:ppsi}(a). It measures four phase-shifted interference fringe images simultaneously, which enables instantaneous and quantitative observation of sound fields. For details of the measurement technique, see, for example, \cite{Ishikawa2016}.

In this experiment, a 12-kHz burst wave generated from a loudspeaker (FOSTEX FT48D) was observed. The sound measured by a microphone placed 20 cm from the loudspeaker is shown in Fig.~\ref{fig:ppsi}(b). The generated sound was a three-cycle 12 kHz burst wave with a peak sound pressure of 13 Pa at the microphone position. The frame rate of the high-speed camera was set to 50 kfps, the number of frames was 1000, the image resolution was 128 $\times$ 128, and the imaging area size was 80 mm $\times$ 80 mm. The optical phase map at each frame was calculated using a typical arctangent operation, followed by 1D unwrapping along the time direction for each pixel. Subsequently, a time-directional \change{high-pass} filter with a cutoff frequency of 500 Hz was applied to remove low-frequency noise components. We call this data the noisy data. The denoising was performed on the noisy data by using the same conventional filters and trained DNNs as in the previous section.

Figure~\ref{fig:ppsi}(c) shows the time-series sound field images of the noisy and denoised data. In the noisy data, random noise and oblique noise patterns appeared in addition to sound waves propagating from the left outside of the image to the right. These oblique patterns should be phase shift errors caused by imperfections in the optical system. 
\change{
Excluding the median filter, it is noticeable that the apparent noise is effectively removed. The difference can be observed in small amplitude wavefronts at 60 $\mu$s and 360 $\mu$s. WFF and Ours (W) exhibit smoother restoration of these small amplitude components compared to other methods.
}
% The Gaussian filter and ST BPF produced smooth wavefronts for the peak wavefront of the burst wave, but noisy components remained in the low-amplitude parts before and after the peak wavefront, as can be seen in the right half of the 60 $\mu$s image and left half of the 240 $\mu$s image. In contrast, Ours (W) restored both the peak and low-amplitude wavefronts smoothly. The wavefront buried in the noise data were also visualized in the results for Ours (W). 
% For Ours (W+S), the restored sound wave amplitudes increased, which is consistent with the numerical results in Fig.~\ref{fig:sim_white}.

\begin{figure}[t]
	\centering
	\includegraphics[width= .9\linewidth]{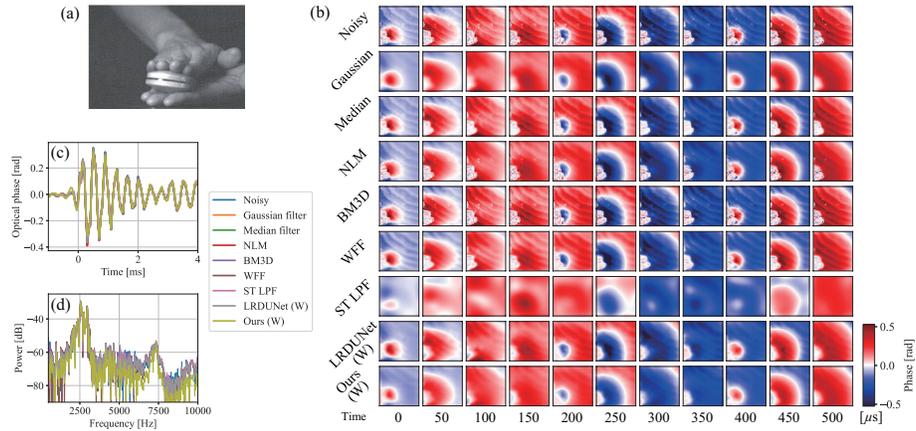}
	\caption{\change{(a) Photo of castanets. (b) Imaging results. (c) and (d) temporal signals and power spectra extracted from the center 4 $\times$ 4 pixels of the images in (b).}}
	\label{fig:ppsi_ca}
\end{figure}

\change{
As a further example of a realistic sound field, we conducted denoising of the sound field radiated from human playing castanets, a type of percussion instrument. Figure~\ref{fig:ppsi_ca}(a) shows a photograph of the castanets. The frame rate of the high-speed camera was set to 20 kfps, the number of frames was 1000, the image resolution was 128 × 128, and the size of the captured area was 100 mm × 100 mm.
}

\change{
Figure~\ref{fig:ppsi_ca}(b) presented the visualization results. A portion of the castanets shadow is visible in the lower-left corner of each image. The waveforms extracted from the image center 4 $\times$ 4 pixels are plotted in Fig.~\ref{fig:ppsi_ca}(c), and the power spectra of the waveforms are shown in Fig.~\ref{fig:ppsi_ca}(d).
Figure~\ref{fig:ppsi_ca}(d) shows that the spectral peak is observed at approximately 2.5 kHz. Since the wavelength of a 2.5 kHz sound wave is approximately 140 mm, only one wavelength or less is visible in the imaging area. It is important to note that, due to this long wavelength, the spatiotemporal low-pass filter (ST LPF) was used instead of the ST BPF. 
The noisy data show the pressure peak (red) and dip (blue) are spread out from the lower left castanets position circularly. Additionally, a noticeable diagonal linear pattern spans across the entire image. It is important to note that this pattern does not indicate sound propagation but spatial noise since it does not propagate through space. The results of the denoising process demonstrate that the Gaussian filter, LRDUet (W), and Ours (W) effectively smooth out this pattern. This means non-moving noise components are removed while propagating sound components remain preserved.
Additionally, as depicted in Fig.~\ref{fig:ppsi_ca}(c), the discrepancies among the temporal waveforms by denoising methods are negligible. Similarly, in Fig.~\ref{fig:ppsi_ca}(d), the first signal component with a peak at 2.5 kHz and the second with a peak at 7.5 kHz remain nearly unaltered in the spectrum. These observations suggest that all denoising methods maintain the temporal and spectral data of the sound field, while specific techniques, such as ours (W), can remove spatial noises. Consequently, we can deduce that the proposed DNN methods are efficient for denoising intricate and practical sound fields.
}

\subsection{Holographic speckle interferometry with Fresnel lens}
\begin{figure}[t]
	\centering
	\includegraphics[width= .7\linewidth]{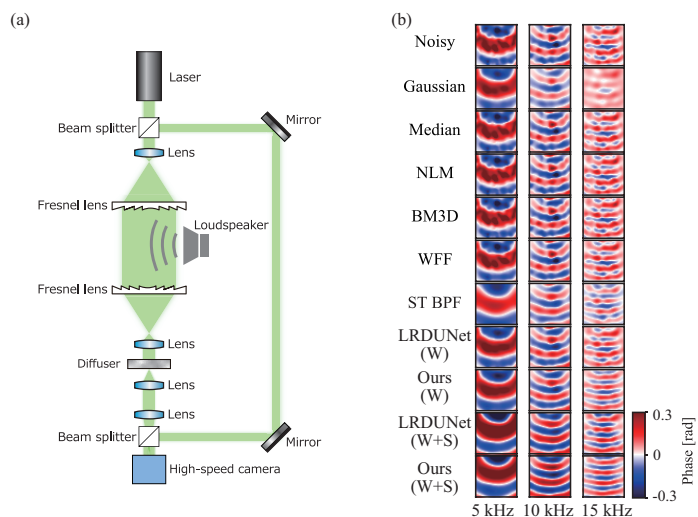}
	\caption{(a) Schematic diagram of the HSI measurement system. The sound field between the two Fresnel lenses is measured. The sSinusoidal waves of 5, 10, and 15 kHz was are emitted from the loudspeaker. (b) Denoising results of harmonic sound fields measured by HSI.}
	\label{fig:hsi}
\end{figure}

An overview of the measurement using HSI is shown in Fig.~\ref{fig:hsi}(a). This experiment used a measurement system with Fresnel lenses, as proposed in \cite{Ishikawa2022}. It was proposed to establishas a lightweight and inexpensive large-aperture sound-field imaging system using Fresnel lenses. However, the measured sound-field images showed significant spatial distortion due to speckle noise. In the original paper, narrow spatial bandpass filters were used for noise reduction, but such narrow filters may not be so useful for practical applications. Here, we investigated the effectiveness of the proposed DNN-based denoising method.

Sinusoidal signals of 5, 10, and 15 kHz were radiated from the same loudspeaker used in the PPSI experiment. The amplitudes were adjusted so that the sound pressure level at the microphone located 20 cm in front of the loudspeaker diaphragm was \change{6.3 Pa (110 dB SPL)} at all frequencies. The frame rate of the high-speed camera was 50 kfps, the number of frames was 1000, the image resolution was 128 $\times$ 128, and the size of the captured area was 100 mm $\times$ 100 mm. The phase maps of the speckle interference fringes were estimated using the 2D FT method~\cite{Goodman2007}, and a complex sound field at each frequency was extracted via 1D FT along the time direction. 

Figure~\ref{fig:hsi}(b) shows the real parts of the noisy and denoised complex amplitudes. 
\change{The noisy data contains the spherical sound wave propagates from top to bottom and low-spatial-frequency wavy patterns that modulate the spherical wavefront. These patterns originate from the recorded specklegrams of this method, as explained in \cite{Ishikawa2022}. For 5 kHz, ST BPF and the four DNNs restored smooth sound waves. For 10 and 15 kHz, one can see that Ours (W+S) provides the smoothest circular wavefronts compared to the other methods.}
Since the same loudspeaker as in the PPSI experiment was used, the harmonic wavefront should be smooth and circular. Therefore, \change{it can be surmised that Ours (W+S) showed the best wavefront restoration performance in speckle sound-field imaging, consistent with the numerical data results.}

\section{Conclusions}
We developed a DNN-based sound-field denoising method in which the trained network decomposes time-varying sound field data into 2D complex amplitude images and denoises each individual image. A 2D sound field simulation with random parameters was used to generate the training dataset. By taking into account the measurement process of the optical system, the network was successfully trained to remove not only white Gaussian noise but also speckle noise. We confirmed that the proposed method was effective on experimental data and that it outperformed conventional denoising methods.

There are questions to be tackled in future work. First, in this study, we employed \change{DnCNN, LRDUNet, and} NAFNet with fixed network sizes. Therefore, the effect of the choice of optical network architecture and its size should be investigated. Second, the simulation method and the number of training data should also be investigated. The generalization abilities against the wavenumber range, complexity of sound fields, and amount and types of noise must depend on the training dataset. Last but not least, it is important to extend the proposed method to different measurement situations, such as spatial 3D data, randomly sampled data, and data with occlusions to provide a versatile denoiser for optically measured sound field data.

% \clearpage
% \appendix
% \section*{Appendix: Speckled sound-field generation}

\begin{backmatter}
%\bmsection{Acknoledgements} 

\bmsection{Disclosures} The authors declare no conflicts of interest.

\bmsection{Data availability} Data underlying the results presented in this paper are not publicly available at this time but may be obtained from the authors upon reasonable request. Code is available on our GitHub repository: \url{https://github.com/nttcslab/deep-sound-field-denoiser}

\bmsection{Supplemental document}
See \href{https://opticapublishing.figshare.com/s/26592e8ab4d2dee2e4eb}{Supplement 1} for supporting content. 

\end{backmatter}

% \bibliography{main}

\end{document}